\documentstyle[epsf,12pt]{article}

  \def\pp{{\mathchoice
          {
              \kern 1pt%
              \raise 1pt
              \vbox{\hrule width5pt height0.4pt depth0pt
                    \kern -2pt
                    \hbox{\kern 2.3pt
                          \vrule width0.4pt height6pt depth0pt
                          }
                    \kern -2pt
                    \hrule width5pt height0.4pt depth0pt}%
                    \kern 1pt
           }
            {
              \kern 1pt%
              \raise 1pt
              \vbox{\hrule width4.3pt height0.4pt depth0pt
                    \kern -1.8pt
                    \hbox{\kern 1.95pt
                          \vrule width0.4pt height5.4pt depth0pt
                          }
                    \kern -1.8pt
                    \hrule width4.3pt height0.4pt depth0pt}%
                    \kern 1pt
            }
            {
              \kern 0.5pt%
              \raise 1pt
              \vbox{\hrule width4.0pt height0.3pt depth0pt
                    \kern -1.9pt  
                    \hbox{\kern 1.85pt
                          \vrule width0.3pt height5.7pt depth0pt
                          }
                    \kern -1.9pt
                    \hrule width4.0pt height0.3pt depth0pt}%
                    \kern 0.5pt
            }
            {
              \kern 0.5pt%
              \raise 1pt
              \vbox{\hrule width3.6pt height0.3pt depth0pt
                    \kern -1.5pt
                    \hbox{\kern 1.65pt
                          \vrule width0.3pt height4.5pt depth0pt
                          }
                    \kern -1.5pt
                    \hrule width3.6pt height0.3pt depth0pt}%
                    \kern 0.5pt
            }
        }}

  \def\mm{{\mathchoice
                       {
                             \kern 1pt
               \raise 1pt    \vbox{\hrule width5pt height0.4pt depth0pt
                                  \kern 2pt
                                  \hrule width5pt height0.4pt depth0pt}
                             \kern 1pt}
                       {
                            \kern 1pt
               \raise 1pt \vbox{\hrule width4.3pt height0.4pt depth0pt
                                  \kern 1.8pt
                                  \hrule width4.3pt height0.4pt depth0pt}
                             \kern 1pt}
                       {
                            \kern 0.5pt
               \raise 1pt
                            \vbox{\hrule width4.0pt height0.3pt depth0pt
                                  \kern 1.9pt
                                  \hrule width4.0pt height0.3pt depth0pt}
                            \kern 1pt}
                       {
                           \kern 0.5pt
             \raise 1pt  \vbox{\hrule width3.6pt height0.3pt depth0pt
                                  \kern 1.5pt
                                  \hrule width3.6pt height0.3pt depth0pt}
                           \kern 0.5pt}
                       }}

\catcode`@=11
\def\un#1{\relax\ifmmode\@@underline#1\else
        $\@@underline{\hbox{#1}}$\relax\fi}
\catcode`@=12

\let\du=\du                     

\def\a{\alpha}
\def\b{\beta}

\def\d{\delta}

\def\f{\phi}
\def\g{\gamma}

\def\j{\psi}

\def\l{\lambda}
\def\m{\mu}

\def\o{\omega}
\def\p{\pi}
\def\q{\theta}

\def\s{\sigma}
\def\t{\tau}

\def\x{\xi}
\def\z{\zeta}
\def\D{\Delta}
\def\F{\Phi}
\def\G{\Gamma}

\def\ve{\varepsilon}
\def\vf{\varphi}

\def\vq{\vartheta}

\def\cf{{\cal F}}

\def\ch{{\cal H}}

\def\bo{{\raise-.5ex\hbox{\large$\Box$}}}               
\def\pa{\partial}                                       
\def\de{\nabla}                                         
\def\pr{\prod}                                          
\def\TH{{\raise.2ex\hbox{$\displaystyle \bigodot$}\mskip-4.7mu \llap H \;}}
\def\face{{\raise.2ex\hbox{$\displaystyle \bigodot$}\mskip-2.2mu \llap {$\ddot
        \smile$}}}                                      

\def\sp#1{{}^{#1}}                              
\def\Tilde#1{\widetilde{#1}}                    
\def\Bar#1{\overline{#1}}                       
\def\VEV#1{\left\langle #1\right\rangle}        
\def\leftrightarrowfill{$\mathsurround=0pt \mathord\leftarrow \mkern-6mu
        \cleaders\hbox{$\mkern-2mu \mathord- \mkern-2mu$}\hfill
        \mkern-6mu \mathord\rightarrow$}
\def\dvec#1{\vbox{\ialign{##\crcr
        \leftrightarrowfill\crcr\noalign{\kern-1pt\nointerlineskip}
        $\hfil\displaystyle{#1}\hfil$\crcr}}}           
\def\dt#1{{\buildrel {\hbox{\LARGE .}} \over {#1}}}     

\def\frac#1#2{{\textstyle{#1\over\vphantom2\smash{\raise.20ex
        \hbox{$\scriptstyle{#2}$}}}}}                   
\def\sfrac#1#2{{\vphantom1\smash{\lower.5ex\hbox{\small$#1$}}\over
        \vphantom1\smash{\raise.4ex\hbox{\small$#2$}}}} 
\def\bfrac#1#2{{\vphantom1\smash{\lower.5ex\hbox{$#1$}}\over
        \vphantom1\smash{\raise.3ex\hbox{$#2$}}}}       
\def\afrac#1#2{{\vphantom1\smash{\lower.5ex\hbox{$#1$}}\over#2}}    

\def\[{\lfloor{\hskip 0.35pt}\!\!\!\lceil}
\def\]{\rfloor{\hskip 0.35pt}\!\!\!\rceil}
\def\Lag{{\cal L}}
\def\du#1#2{_{#1}{}^{#2}}

\def\tr{{\rm tr}}
\def\Tr{{\rm Tr}}

\def\un{\underline}
\def\fracmm#1#2{{{#1}\over{#2}}}

\def\low#1{{\raise -3pt\hbox{${\hskip 0.75pt}\!_{#1}$}}}

\def\Dot#1{\buildrel{_{_{\hskip 0.01in}\bullet}}\over{#1}}
\def\dt#1{\Dot{#1}}

\def\Tilde#1{{\widetilde{#1}}\hskip 0.015in}

\def\sbar#1{\stackrel{*}{\Bar{#1}}}

\newskip\humongous \humongous=0pt plus 1000pt minus 1000pt
\def\caja{\mathsurround=0pt}
\def\eqalign#1{\,\vcenter{\openup2\jot \caja
        \ialign{\strut \hfil$\displaystyle{##}$&$
        \displaystyle{{}##}$\hfil\crcr#1\crcr}}\,}
\newif\ifdtup

\def\ref#1{$\sp{#1)}$}

\def\pl#1#2#3{Phys.~Lett.~{\bf {#1}B} (19{#2}) #3}
\def\np#1#2#3{Nucl.~Phys.~{\bf B{#1}} (19{#2}) #3}

\def\pr#1#2#3{Phys.~Rev.~{\bf D{#1}} (19{#2}) #3}
\def\cqg#1#2#3{Class.~and Quantum Grav.~{\bf {#1}} (19{#2}) #3}

\parskip=\medskipamount                 
\thicklines                         

\begin{document}


\thispagestyle{empty}               

\def\border{                                            
        \setlength{\unitlength}{1mm}
        \newcount\xco
        \newcount\yco
        \xco=-24
        \yco=12
        \begin{picture}(140,0)
        \put(-20,11){\tiny Institut f\"ur Theoretische Physik Universit\"at
Hannover~~ Institut f\"ur Theoretische Physik Universit\"at Hannover~~
Institut f\"ur Theoretische Physik Hannover}
        \put(-20,-241.5){\tiny Institut f\"ur Theoretische Physik Universit\"at
Hannover~~ Institut f\"ur Theoretische Physik Universit\"at Hannover~~
Institut f\"ur Theoretische Physik Hannover}
        \end{picture}
        \par\vskip-8mm}

\def\headpic{                                           
        \indent
        \setlength{\unitlength}{.8mm}
        \thinlines
        \par
        \begin{picture}(29,16)
        \put(75,16){\line(1,0){4}}
        \put(80,16){\line(1,0){4}}
        \put(85,16){\line(1,0){4}}
        \put(92,16){\line(1,0){4}}

        \put(85,0){\line(1,0){4}}
        \put(89,8){\line(1,0){3}}
        \put(92,0){\line(1,0){4}}

        \put(85,0){\line(0,1){16}}
        \put(96,0){\line(0,1){16}}
        \put(92,16){\line(1,0){4}}

        \put(85,0){\line(1,0){4}}
        \put(89,8){\line(1,0){3}}
        \put(92,0){\line(1,0){4}}

        \put(85,0){\line(0,1){16}}
        \put(96,0){\line(0,1){16}}
        \put(79,0){\line(0,1){16}}
        \put(80,0){\line(0,1){16}}
        \put(89,0){\line(0,1){16}}
        \put(92,0){\line(0,1){16}}
        \put(79,16){\oval(8,32)[bl]}
        \put(80,16){\oval(8,32)[br]}

        \end{picture}
        \par\vskip-6.5mm
        \thicklines}

\border\headpic {\hbox to\hsize{
\vbox{\noindent DESY 97 -- 007  \hfill January 1997 \\
ITP--UH--03/97 \hfill hep-th/9701158 }}}

\noindent
\vskip1.3cm
\begin{center}

{\Large\bf The effective hyper-K\"ahler potential  
\vglue.1in   in the ~N=2~ supersymmetric QCD~\footnote{Supported 
in part by the `Deutsche Forschungsgemeinschaft' 
and the NATO grant CRG 930789}}
\vglue.3in

Sergei V. Ketov \footnote{
On leave of absence from:
High Current Electronics Institute of the Russian Academy of Sciences,
\newline ${~~~~~}$ Siberian Branch, Akademichesky~4, Tomsk 634055, Russia}

{\it Institut f\"ur Theoretische Physik, Universit\"at Hannover}\\
{\it Appelstra\ss{}e 2, 30167 Hannover, Germany}\\
{\sl ketov@itp.uni-hannover.de}
\end{center}
\vglue.2in
\begin{center}
{\Large\bf Abstract}
\end{center}

The effective low-energy hyper-K\"ahler potential for a massive N=2 matter in
N=2 super-QCD is investigated. The N=2 extended supersymmetry severely 
restricts that N=2 matter self-couplings so that their {\it exact} form can be
fixed by a few parameters, which is apparent in the N=2 harmonic superspace.
In the N=2 QED with a single matter hypermultiplet, the one-loop perturbative 
calculations lead to the {\it Taub-NUT} hyper-K\"ahler metric in the massive 
case, and a free metric in the massless case. It is remarkable that the naive 
non-renormalization `theorem' does not apply. There exists a manifestly N=2 
supersymmetric duality transformation converting the low-energy effective 
action for the N=2 QED hypermultiplet into a sum of the quadratic and the
improved (non-polynomial) actions for an N=2 tensor multiplet. The duality
transformation also gives a simple connection between the low-energy effective
action in the N=2 harmonic superspace and the component results.

\newpage

\section{Introduction} 

The effective action has proved to be very useful in quantum field theory. The 
full effective action is however non-local and highly complicated, and it is 
usually impossible to calculate it. Amongst the practical approximations capable
to go beyond the standard (loop) perturbation theory, an expansion of the 
effective action in momenta (or in the number of spacetime derivatives) plays a 
prominent role. Its leading term --- the so-called {\it low-energy effective 
action} (LEEA) -- encodes important information about the spectrum and static 
couplings in the full quantum theory.
 
In the famous papers \cite{sw1,sw2} Seiberg and Witten described a construction
of the {\it exact} (i.e. perturbative and non-perturbative) LEEA for N=2 
supersymmetric Yang-Mills theories, and subsequently generalised it to the case 
of N=2 super-QCD, i.e. in a presence of N=2 matter. In N=2 superspace, the 
four-dimensional N=2 super-Yang-Mills (SYM) theory is described by the Lie 
algebra-valued (reduced) chiral N=2 superfield strength $W$, which is sometimes 
called N=2 superpotential.~\footnote{The term `superpotential' is somewhat 
confusing here, since the $W$ is a constrained superfield, \newline ${~~~~~}$ 
unlike the unconstrained N=2 SYM pre-potential $V^{++}$ to be introduced in 
sect.~2.} The most general {\it Ansatz} for the (Wilsonian) LEEA of 
spontaneously broken (abelian) N=2 SYM theory takes the form of a chiral N=2 
superspace integral~\cite{gates}
$$ S_{\cf}=\fracmm{1}{4\p}{\rm Im}\,\Tr\int d^4xd^4\q \,\cf(W) 
\eqno(1.1)$$
to be determined by a single holomorphic function $\cf(W)$. Demanding 
renormalizability of the initial (microscopic) N=2 SYM action requires the 
function $\cf$ to be quadratic in $W$. It was shown by Seiberg and Witten 
\cite{sw1,sw2} how to determine the abelian LEEA function $\cf$ exactly, 
provided that N=2 supersymmetry is not dynamically broken. As regards the 
{\it non-abelian} LEEA, it also receives contributions from the {\it 
non-holomorphic} coupling
$$ S_{\ch}=\int d^4xd^4\q d^4\bar{\q}\,\ch(W,\bar{W})~,\eqno(1.2)$$
as was demonstrated in refs.~\cite{wgr,piwest}. The real function $\ch$ was 
(partially) fixed in ref.~\cite{wgr}.

The N=2 supersymmetric matter is described by hypermultiplets~\cite{fs}. 
Each N=2 hypermultiplet contains a pair of complex scalars and a Dirac spinor, 
all transforming in the same gauge group representation, which can be different 
from the adjoint representation. By N=2 supersymmetry, the hypermultiplet 
scalars parametrise a hyper-K\"ahler manifold in the LEEA. It is therefore of
interest to determine the associated hyper-K\"ahler metric or the corresponding
effective hyper-K\"ahler potential for that metric. In order to write down the
N=2 superspace Ansatz, which would be analogous to eq.~(1.1) but now for the N=2
matter, one needs a general manifestly N=2 supersymmetric off-shell formulation 
for the hypermultiplets. Such a formulation is known, and it is provided by the 
N=2 {\it harmonic superspace} (HSS)~\cite{gikos}. Both the conventional N=2 
superspace methods and the N=1 superspace approach are not adequate for 
constructing the hypermultiplet LEEA, since either they have problems with a 
gauge-fixing and ghosts-for-ghosts (e.g., as in ref.~\cite{hst}), or they suffer
from the absence of manifest N=2 supersymmetry (e.g., as in 
refs.~\cite{piwest,gru}), which do not even allow one to write down the proper
Ansatz for the N=2 matter LEEA.
 
In this Letter, I consider the N=2 super-QCD with the gauge group 
$SU(N_{\rm c})$ and the N=2 matter to be represented by $N_{\rm f}$ charged 
massive hypermultiplets, each transforming in the fundamental representation 
$\underline{N_{\rm c}}$ of the gauge group. In sect.~2 the microscopic  
(renormalizable) N=2 supersymmetric gauge theory is formulated in the N=2 HSS. 
In sect.~3, I introduce the most general Ansatz for the hypermultiplet LEEA, 
which determines the form of the effective hyper-K\"ahler potential  up to a 
few (finite) parameters whose actual appearance is connected to the (chiral)
symmetry breaking. It is demonstrated in sect.~4 that the proposed terms in the
N=2 matter LEEA do actually appear in the one-loop perturbation theory, by 
using the HSS Feynman rules of refs.~\cite{hss,ohta}. In sect.5 a duality 
(Legendre) transformation is used to relate the simplest non-trivial N=2 matter
LEEA to be written in the HSS for the N=2 QED, to the equivalent N=2 matter 
action in the projective N=2 superspace, which also provides a simple connection
to the component results. The conclusions are summarized in sect.~6.

\section{The setup}

In the HSS formalism, the standard N=2 superspace $(x^m,x^5,\q^{\a}_i,
\bar{\q}^{\dt{\a}i})$, $m=0,1,2,3$; $\a=1,2$, and $i=1,2$, is extended by 
adding the bosonic variables (or `zweibeins') $u^{\pm i}$ parametrizing the
sphere $S^2\sim SU(2)/U(1)$~:~\footnote{I use the notation and conventions as 
in ref.~\cite{gikos}. In particular, the $SU(2)$ indices are raised and \newline
${~~~~~}$ lowered with the antisymmetric Levi-Civita symbols $\ve_{ij}$ and  
$\ve^{ij}$, $\ve^{12}=-\ve_{12}=1$. The ordinary \newline ${~~~~~}$ complex 
conjugation is detoned by bar. The extra bosonic coordinate $x^5$ is needed to 
take into \newline ${~~~~~}$ account central charges~\cite{fs}.}
$$ \left( \begin{array}{c} u^{+i} \\ u^{-i}\end{array}\right) \in SU(2)~,
\quad {\rm so~~that}\quad u^{+i}u^-_i=1~,\quad
 u^{+i}u^+_i=u^{-i}u^-_i=0~;\quad i=1,2~.\eqno(2.1)$$

Instead of using an explicit parametrization for the sphere, it is more 
convenient to use the functions of zweibeins but consider only those of them
which carry a definite $U(1)$ charge $q$ to be defined by $q(u^{\pm}_i)=\pm 1$. 
It leads to the simple integration rules~\cite{gikos}
$$ \int du =1~,\qquad \int du\, u^{+(i_1}\cdots u^{+i_m}u^{-j_1}\cdots
u^{-j_n)}=0~,\quad {\rm when}\quad m+n>0~.\eqno(2.2)$$
It follows that the integral over {\it any}~ $U(1)$-charged quantity vanishes.  

In addition to the usual complex conjugation, there exists a star conjugation 
that only acts on the $U(1)$ indices, $(u^+_i)^*=u^-_i$ and
$(u^-_i)^*=-u^+_i$. Accordingly, one has~\cite{gikos}
$$ \sbar{u^{\pm i}}=-u^{\pm}_i~,\qquad  \sbar{u^{\pm}_i}=u^{\pm i}~.\eqno(2.3)$$

One can also introduce the covariant derivatives  (in the central basis)
with respect to the zweibeins, which preserve the defining conditions (2.1),
namely,
$$ D^{++}_{\rm c}=u^{+i}\fracmm{\pa}{\pa u^{-i}}~,\quad 
D^{--}_{\rm c}=u^{-i}\fracmm{\pa}{\pa u^{+i}}~,\quad 
D^{0}_{\rm c}=u^{+i}\fracmm{\pa}{\pa u^{+i}}-u^{-i}\fracmm{\pa}{\pa u^{-i}}~.
\eqno(2.4)$$

A key feature of the N=2 HSS is an existence of the {\it analytic}
subspace parametrised by the coordinates
$$ (\z,u)=\left\{ \begin{array}{c}
x^m_{\rm analytic}=x^m-2i\q^{(i}\s^m\bar{\q}^{j)}u^+_iu^-_j~, \quad 
x^5_{\rm analytic}=x^5+i(\q^+\q^--\bar{\q}^+\bar{\q}^-)~,\\
\q^+_{\a}=\q^i_{\a}u^+_i~,\quad \bar{\q}^+_{\dt{\a}}=\bar{\q}^i_{\dt{\a}}u^+_i~,
\quad u^{\pm}_i \end{array} \right\}~.\eqno(2.5)$$
It is invariant under N=2 supersymmetry, and is closed under the combined 
conjugation of eq.~(2.3) \cite{gikos}. That allows one to define the 
{\it analytic} superfields of any $U(1)$ charge $q$, by the analyticity 
conditions
$$D^+_{\a}\f^{(q)}=\bar{D}^+_{\dt{\a}}\f^{(q)}=0~,\quad {\rm where}\quad
D^+_{\a}=D^i_{\a}u^+_i \quad {\rm and}\quad
\bar{D}^+_{\dt{\a}}=\bar{D}^i_{\dt{\a}}u^+_i~,\eqno(2.6)$$
and introduce the analytic measure $d\z^{(-4)}du\equiv (d^4xdx^5)_{\rm analytic}
d^2\q^+d^2\bar{\q}^+du$ of charge $(-4)$ as well, so that the full measure in the
N=2 HSS can be written down as
$$ d^4xdx^5d^4\q d^4\bar{\q}du=d\z^{(-4)}du(D^+)^4~,\quad {\rm where}\quad
(D^+)^4=\fracmm{1}{16}(D^{+\a}D_{\a}^+)(\bar{D}^{+}_{\dt{\a}}\bar{D}^{+\dt{\a}}
)~.\eqno(2.7)$$
In the analytic subspace, the harmonic derivative 
$D^{++}=D^{++}_{\rm c}-2i\q^+\s^m\bar{\q}^+\pa_m$ obviously preserves 
analyticity, and it allows one to integrate by parts.

An action for the most general renormalizable N=2 supersymmetric gauge theory 
with matter in the N=2 HSS is given by
$$ S =\fracmm{1}{4\p T(R)}\,{\rm Im}\,\Tr \int d^4xd^4\q dx^5du\,\fracmm{1}{2}
\t W^2 + 
\int d\z^{(-4)}du\,\tr\,\sbar{\f}{}^+(D^{++}+iV^{++})\f^+~,\eqno(2.8)$$
where the coupling constant is $\t\equiv\fracmm{\q}{2\p}+\fracmm{4\p i}{g^2}$ as
usual, and the integration over $x^5$ is defined similarly to that in eq.~(2.2),
namely,
$$ \int dx^5\,=1~,\qquad \int dx^5\,e^{imx^5}=0~,\quad {\rm when}\quad
m\neq 0~.\eqno(2.9)$$
 The chiral gauge N=2 superfield strength $W$ is $u$-independent and 
$x^5$-independent as in sect.~1, and it has to be considered as a (complicated) 
function of an {\it unconstrained} and {\it  analytic} Lie algebra-valued gauge 
superfield $V^{++}$. The $V^{++}$ enters as an extention for the 
$D^{++}$-connection, satisfies a reality condition $\Bar{V^{++}}^{\,*}=V^{++}$,
and has no dependence upon $x^5$ too. The explicit formula $W(V^{++})$ is given
in refs.~\cite{gikos,hss}, and it is not needed for our purposes. Each of the 
hypermultiplets representing N=2 matter in eq.~(2.8) is described in the N=2 HSS
by a {\it complex analytic} superfield $\f^+$ transforming in a representation 
$R$ of the gauge group with generators $T^a$,~\footnote{The superfield $V^{++}$
in the second term of eq.~(2.8) is also valued in the R-representation.}
 $\tr(T^aT^b)=T(R)\d^{ab}$.  In the N=2 super-QCD, the gauge group is 
$SU(N_{\rm c})$, and the representation $R$ is
a reducible combination of $N_{\rm f}$ fundamental representations. Unlike the 
$V^{++}$, each superfield $\f^+$ is assumed to be $x^5$-dependent as 
$\exp(im_Ax^5)$, thus describing a massive hypermultiplet of mass $m_A$. As is 
well known, the mass of a hypermultiplet can only come from the central charges 
in the N=2 superalgebra since, otherwise, the number of the massive 
hypermultiplet components has to be increased.  

The gauge-invariant HSS action (2.8) has to be supplemented by a gauge-fixing 
term and the FP ghost term. For HSS perturbative calculations in the 
background-field method, the N=2 supersymmetric Feynman gauge is convenient since
the gauge-fixed kinetic term (without a $\q$-term) for the $V^{++}$ superfield
is particularly simple,
$$ S_{V-{\rm kin.}}=\fracmm{1}{4T(R)g^2}\,\Tr\,\int d\z^{(-4)}du\,V^{++}\bo
V^{++}~.\eqno(2.10)$$
The full list of the HSS Feynman rules can be found in refs.~\cite{hss,ohta}
(see also sect.~4).

\section{The effective hyper-K\"ahler potential \\ 
         and symmetry breaking in N=2 QCD}

Let's first summarize the classical symmetries of the microscopic N=2 super-QCD
action (2.8). The theory has the local $SU(N_{\rm c})$ gauge symmetry, the global
N=2 supersymmetry, a global $SU(2)_{\rm A}$ symmetry which is an automorphism of
the N=2 superalgebra (it rotates its two supercharges), and an R-symmetry 
$U(1)_{\rm R}$. If all the hypermultiplets have the same mass $m$, the theory
has an $SU(N_{\rm f})$ flavour symmetry which is broken down to a smaller
subgroup when the masses are not equal. If {\it all} the masses are different,
the $SU(N_{\rm f})$ is broken down to $U(1)^{N_{\rm f}}$.

In quantum theory some of the above symmetries may be broken or become anomalous.
Based on calculations of the Witten index~\cite{witten}, one can argue that N=2
supersymmetry should remain unbroken in the N=2 QCD. Accordingly, the global
$SU(2)_{\rm A}$ symmetry should also be a symmetry of the quantum theory
\cite{sw1,sw2}. The R-symmetry is well-known to be anomalous due to the
non-perturbative instanton effects, so that it is actually broken to a discrete 
subgroup. The (spontaneous) gauge symmetry breaking is dependent upon the vacuum
which, in its turn, can be found by minimizing the component (tree) scalar 
potential. The vacuum solutions with unbroken sypersymmetry are 
known~\cite{sbr}. In particular, if the hypermultiplet masses do not vanish, one
has $\VEV{q^i}=0$, where $\left. q^i\equiv\f^i\right|$ is the leading scalar 
component of the on-shell hypermultiplet, $\f^+=\f^i(x,\q,\bar{\q})u^+_i$, so 
that only the leading scalar component $\left. A\equiv W\right|$ of the vector 
N=2 multiplet can have a non-vanishing vacuum expectation value (a Coulomb 
phase). If all the hypermultiplet masses are zero, there exist solutions with 
$\VEV{q^i}\neq 0$ (the scalar potential has flat directions) but $\VEV{A}=0$ 
(see ref.~\cite{sbr} for details) -- it is a Higgs phase, according to the 
classification of refs.~\cite{sw1,sw2}. In the Higgs phase, there are no
monopoles or dyons, and there should be, therefore, no non-perturbative 
corrections to the hyper-K\"ahler potential. In the Coulomb phase, the abelian 
gauge symmetry remains unbroken, while all the hypermultipets are massive, and 
one expects the one-loop hyper-K\"ahler potential to be exact, since there are 
no instantons ~\cite{sw2}.   

We are now in a position to consider the LEEA Ansatz for the effective 
hypermultiplet self-couplings in the N=2 HSS. On dimensional grounds, it has to
be an integral of a local quantity over the analytic subspace. Since the local
quantity has to (i) be constructed out of $\f^+$, $\sbar{\f}{}^+$ and $D^{++}$, 
(ii) have the $U(1)$ charge $q=+4$, and (iii) preserve the $SU(2)_{\rm A}$ 
global invariance, the most general Ansatz appears to be fixed up to a finite 
number of parameters $(\l,\b,\g)$ as follows ({\it cf.} ref.~\cite{hss}):
$$ S_{\rm hyper-K.}=\fracmm{1}{2}\int d\z^{(-4)}du\,\left( \sbar{\f}{}^+
\dvec{D}{}^{++}\f^+ + \Lag_{\rm int.}^{(+4)}\right)~,\eqno(3.1)$$
where
$$ \Lag_{\rm int.}^{(+4)}=\fracmm{\l}{2}(\sbar{\f}{}^+)^2(\f^+)^2 +
\left[ \b \sbar{\f}{}^+(\f^+)^3 + \g(\f^+)^4 + {\rm h.c.} \right]~.\eqno(3.2)$$
In the case of N=2 QCD with the charged massive hypermultiplets in the 
fundamental (complex) representation of the gauge group (in the microscopic
action), the unbroken (abelian) symmetries of the quantum theory still require
$\b=\g=0$. Hence, we are left with the quartic self-interactions
$$ \Lag_{\rm int.}^{{\rm QCD}(+4)}= \sum^{N_{\rm f}}_{{\rm flavour}
\atop (ABCD)}  \sum^{N_{\rm c}}_{{\rm colour}\atop (abcd)}\l^{ABCD}_{abcd}
\sbar{\f}{}^+_{Aa}
\f^+_{Bb}\sbar{\f}{}^+_{Cc}\f^+_{Dd}~,\eqno(3.3)$$
whose particular structure (i.e. the non-vanishing $\l$'s) is dependent upon the
chiral and flavour symmetry breaking under consideration.

In the case of the abelian N=2 super-QED with a single complex hypermultiplet, 
eq.~(3.3) has only one term,
$$ \Lag_{\rm int.}^{{\rm QED}(+4)}=\fracmm{\l}{2}(\f^+)^2(\sbar{\f}{}^+)^2~.
\eqno(3.4)$$
It is worth mentioning that the constant $\l$ has dimension $\[m\]^{-2}$ in units
of mass. It is also clear that the whole Ansatz (3.1) is in conflict with the 
naive non-renormalization `theorem' in the ordinary superspace~\cite{grs} and in
the HSS~\cite{hss,ohta}, that formally forbids quantum contributions of the form
of an integral over a chiral or an analytic subspace of superspace. Hence, it
has to be explained how the quantum (finite) analytic corrections to the 
effective hyper-K\"ahler potential are nevertheless possible (see the next 
sect.~4).

\section{The one-loop effective hyper-K\"ahler \\
         potential in the N=2 super-QED}

The one-loop local contribution to the effective hyper-K\"ahler potential in the
N=2 super-QED can be easily calculated by using the HSS Feynman rules for the 
theory (2.8) in the N=2 super-Feynman gauge~\cite{hss,ohta}. One expands the
action to the second order around a hypermultiplet background $\F^+$,
$\f^+=\F^+ + \vf^+$. The propagators for the quantum HSS fields 
$\vf^+$ and $V^{++}$ are defined by the kinetic terms, 
as in refs.~\cite{hss,ohta}:
$$\VEV{\sbar{\vf}{}^+(p_1,\q_1,u_1)\vf^+(p_2,\q_2,u_2)}=
\fracmm{i}{(p_2-p_1)^2+m^2}\, \fracmm{(D_1^+)^4(D_2^+)^4}{(u_1^{+i}u_{2i}^+)^3}
\,\d^8(\q_1-\q_2)~,\eqno(4.1)$$
and (after rescaling $V$ by the factor of $g$)
$$ \VEV{V^{++}(p_1,\q_1,u_1)V^{++}(p_2,\q_2,u_2)}=
\fracmm{i}{(p_2-p_1)^2}(D_1^+)^4\d^8(\q_1-\q_2)\d^{(-2,2)}(u_1,u_2)~,\eqno(4.2)$$
where $\d^{(-2,2)}(u_1,u_2)$ is one of the harmonic delta-functions on $S^2$,
defined by the equations
$$ \int dv\,\d^{(q,-q)}(u,v)f^{(p)}(v)=\d^{pq}f^{(q)}(u) \eqno(4.3)$$
for any regular function $f^{(p)}(u)$ of the $U(1)$ charge $(p)$. The explicit 
form of $\d^{(q,-q)}(u,v)$ can be found in ref.~\cite{hss}, but it is not really
needed for perturbative calculations.

Since the FP ghosts do not couple to the hypermultiplet, they can be ignored at 
one loop. Then the only relevant vertices are $\sbar{\F}(p_1)V(k)\vf(p_2)$
and $\sbar{\vf}(p_1)V(k)\F(p_2)$, each contributing $-g(2\p)^4\d(p_1-p_2-k)$ to
the Feynman rules, like that in the ordinary QED (the momentum integration is
implied). It leaves us with only one HSS graph $\G_4$ that has to be calculated 
(Fig.~1).

\begin{figure}[b]
\vglue.1in
\makebox{
\epsfxsize=4in
\epsfbox{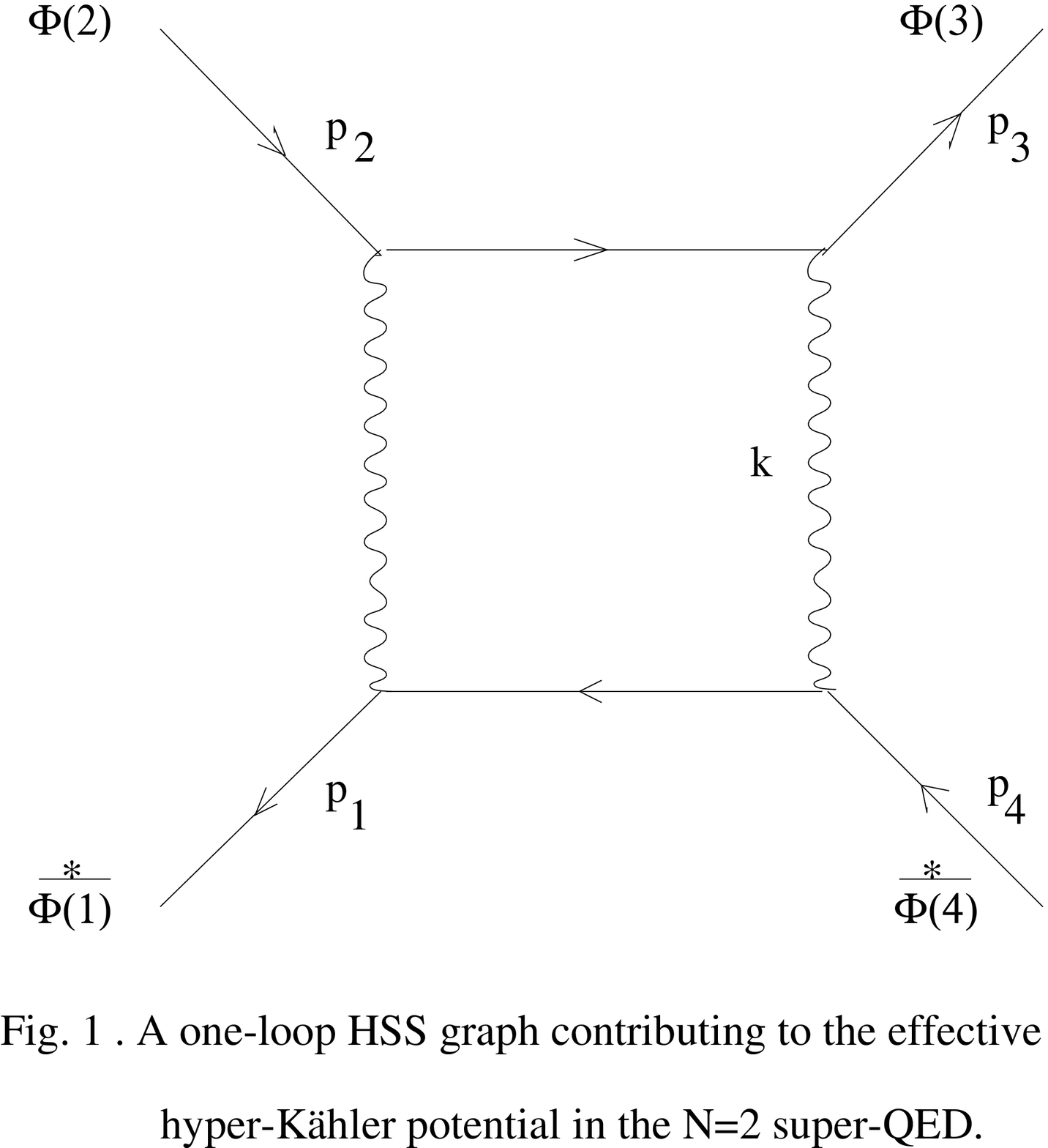}
}
\end{figure}

First, one does all the $\q$-integrations but one, by taking the factor 
$(D_1^+)^4(D_2^+)^4$ off a hypermultiplet propagator and using the 
identity~\cite{hss}
$$ \d^8(\q_1-\q_2)(D_1^+)^4(D_2^+)^4\d^8(\q_1-\q_2)=(u^{+i}_1u^+_{2i})^4
\d^8(\q_1-\q_2)~.\eqno(4.4)$$
As a result, the {\it full} N=2 HSS Grassmann measure $d^8\q$ is restored, just 
in agreement with the non-renormalization theorem ({\it cf.} 
refs.~\cite{hss,ohta}),
$$\eqalign{
\G_4=~&~ g^4\int \fracmm{d^4p_1d^4p_2d^4p_3d^4p_4}{(2\p)^{16}}
\int \fracmm{du_1du_2}{(u^{+i}_1u^+_{2i})^2} \int d^8\q 
\sbar{\F}{}^+(1)\F^+(2)\F^+(3)\sbar{\F}{}^+(4) \times \cr
~&~\times \int 
\fracmm{d^4k \, \d(p_1-p_2+p_3-p_4)}{k^2(k+p_1-p_4)^2[(k-p_3)^2+m^2]
[(k-p_4)^2+m^2]}~.\cr}\eqno(4.5)$$
It is worth mentioning that the loop momentum integral is UV-convergent.

Since the external legs are on-shell, $D^{++}\F^+=D^{++}\sbar{\F}{}^+=0$,
the identities~\cite{hss}
$$\F^+(u)= D^{++}D^{--} \F^+(u) \quad {\rm and}\quad D_1^{++}
\fracmm{1}{(u^{+i}_1u^+_{2i})^2}= D^{--}_1\d^{(2,-2)}(u_1,u_2)\eqno(4.6)$$
can be used to eliminate one of the $u$-integrations in the low-energy 
approximation. The remaining harmonic superspace integral has just two 
$D^{--}$-insertions, and it leads to the appearance of an {\it analytic} term
 because of eq.~(2.7) and another indentity~\cite{hss}
$$ -\,\fracmm{1}{2}(D_1^+)^4(D^{--})^2\F^+(u)=\,\bo\,\F^+(u)=m^2\F^+(u)~.
\eqno(4.7)$$
The low-energy analytic contribution takes the form
$$\G_4=g^4\int \fracmm{d^4k}{(2\p)^4}\,\fracmm{m^2}{k^4(k^2+m^2)^2}
\int d\z^{(-4)}du (\sbar{\F}{}^+)^2(\F^+)^2~, \eqno(4.8)$$
where the loop integral has to be regularised by restricting $k^2\geq \m^2$,
with $\m$ being an IR (Wilsonian) cutoff. One easily finds that
$$\l=\fracmm{g^4}{8\p^2\m^2}f(y)~,~{\rm where}~~f(y)=\fracmm{\ln(1+y)}{y}
-\fracmm{1}{1+y}~,~{\rm with}~~y\equiv\fracmm{m^2}{\m^2}~~{\rm and}~
f(0)=0~.\eqno(4.9)$$

The loophole in the non-renormalization `theorem', which is apparent above,
seems to be quite similar to the other counter-examples found earlier in 
ref.~\cite{west} for some supersymmetric theories with massless fields. 

\section{ The effective hyper-K\"ahler metric and \\
          a duality transformation for N=2 QED}

Given the hypermultiplet LEEA in the form (3.1) with the interaction (3.4), 
it is still non-trivial to extract the explicit hyper-K\"ahler metric of the
associated {\it non-linear sigma-model} (NLSM) which arises after eliminating
the infinite tower of auxiliary fields contained in the harmonic expansion of
the HSS superfield $\f(\z,u)$ via equations of motion. Fortunately, just in the
case under consideration, it was already done in ref.~\cite{hkme} by solving
constraints appearing in the full HSS equation of motion
$$ D^{++}\f^+ +\l(\sbar{\f}{}^+\f^+)\f^+=0~.\eqno(5.1)$$
The remaining `true' equation in components has the form of a NLSM equation of 
motion, whose metric can be written down in proper (`spherical') coordinates as 
the {\it Taub-NUT} metric~\cite{taub}
$$ ds^2=\fracmm{1}{2}\,\fracmm{r+M}{r-M}dr^2 + \fracmm{1}{2}(r^2-M^2)
(d\vq^2+\sin^2\vq d\vf^2) +2M^2\,\fracmm{r-M}{r+M}(d\j +\cos\vq d\vf)^2~,
\eqno(5.2)$$
where $M\equiv \fracmm{1}{2}\l^{-1/2}\sim g^{-2}$ is the mass of the Taub-NUT 
gravitational instanton.

It is worth mentioning that the solitonic mass $M$ is proportional to the
{\it inverse} gauge coupling constant squared, so that its origin is
{\it non}-perturbative with respect to the microscopic action (2.8). That is
because its derivation involves a kind of duality transformation relating weak 
and strong couplings, similarly to that considered in refs.~\cite{sw1,sw2} 
(see e.g., ref.~\cite{krev} for a review).

There exists a manifestly N=2 supersymmetric duality (Legendre) transformation
in the N=2 HSS~\cite{npd} that transforms the hypermultiplet action
$$ S_{\rm hyper.}= \fracmm{1}{2}\int d\z^{(-4)}du\,\left[ \sbar{\f}{}^+
\dvec{D}{}^{++}\f^+ + \fracmm{\l}{2}(\f^+)^2(\sbar{\f}{}^+)^2\right]
\eqno(5.3)$$
into the dual one to be written in terms of {\it dimensionless} real analytic 
HSS superfields $L^{++}$ and $\o$,
$$ S_{\rm dual}=\l S_{\rm free} + S_{\rm improved}~,\eqno(5.4)$$
where
$$  S_{\rm free}=\fracmm{\m^4}{2}\int d\z^{(-4)}du\,\left[
(L^{++})^2 + \o D^{++}L^{++}\right] \eqno(5.5)$$
describes a (non-conformal) free N=2 tensor multiplet in the
N=2 HSS,~\footnote{An equivalence to the standard N=2 superspace formulation of 
the N=2 tensor multiplet, given \newline ${~~~~~}$ by the constraints
$D_{\a}{}^{(i}L^{jk)}=\bar{D}_{\dt{\a}}{}^{(i}L^{jk)}=0$, can be easily 
established one-shell~: a variation of \newline ${~~~~~}$ the action (5.5) with 
respect to the Lagrange multiplier $\o$ yields a HSS constraint $D^{++}L^{++}=0$
\newline ${~~~~~}$ which implies $L^{++}(\z,u)=u^+_iu^+_jL^{ij}$ as well as 
the defining constraints for the $L^{ij}(x,\q,\bar{\q})$.} 
whereas
$$ S_{\rm improved} = \fracmm{\m^2}{2}\int d\z^{(-4)}du\,\left[
(g^{++})^2 + \o D^{++}L^{++}\right]~, \eqno(5.6)$$
with~\cite{npd}
$$ g^{++}(L,u)\equiv \fracmm{2(L^{++}-2iu_1^+u_2^+)}{1+\sqrt{
1-4u_1^+u_2^+u_1^-u_2^--2iL^{++}u_1^-u_2^-}}~,\eqno(5.7)$$
describes the {\it improved} (i.e. N=2 superconformally invariant) 
action~\cite{holl} for the same N=2 tensor multiplet. The duality transformation
reads as follows~\cite{npd}:
$$ \eqalign{
\f^+\,=~&-i(2u^+_1 +ig^{++}u^-_1)e^{-i\o/2}~,\cr
\sbar{\f}{}^+\,=~&+i(2u^+_2 -ig^{++}u^-_2)e^{+i\o/2}~,\cr}
\eqno(5.8)$$
and implies $\sbar{\f}{}^+\f^+=2iL^{++}$ in particular. Therefore, the N=2 HSS
action (5.3) is dual to a sum of the naive (quadratic) and improved 
(non-polynomial) actions for an N=2 tensor multiplet. Both actions are known
both in components~\cite{holl} and in terms of the ordinary N=1 
superfields~\cite{rocek}. The most elegant formulation with a finite number of
auxiliary fields exists in the {\it projective} N=2 superspace where a single
complex $CP(1)$ coordinate $\x$ plays the role of the HSS zweibeins,
$u_i\to \x_i=(1,\x)$. The defining N=2 tensor multiplet constraints in the 
standard N=2 superspace (see the footnote \# 6) imply
$$\de_{\a}G~\equiv~(D^1_{\a}+\x D^2_{\a})G=0 \quad {\rm and}\quad
\D_{\dt{\a}}G~\equiv~(\bar{D}^1_{\dt{\a}}+\x\bar{D}^2_{\dt{\a}})G=0 \eqno(5.9)$$
fior {\it any} function $G(Q(\x),\x)$ with $Q(\x)\equiv\x_i\x_jL^{ij}$. Hence,
after integrating the function $G$ over the rest of the standard N=2 superspace
coordinates, one gets an N=2 superinvariant~\cite{sweden,ketov}
$$\int d^4 x \,\fracmm{1}{2\p i}\oint_C d\x\,\Tilde{\de}^2\Tilde{\D}^2G(Q,\x)~,
\eqno(5.10)$$
where $\Tilde{\de}_{\a}~\equiv~\x D^1_{\a}- D^2_{\a}$ and 
$\Tilde{\D}_{\dt{\a}}~\equiv~\x \bar{D}^1_{\dt{\a}}-\bar{D}^2_{\dt{\a}}$. The
sum of the naive and improved N=2 tensor multiplet actions in the projective 
N=2 superspace is given by~\cite{sweden}
$$ S_{dual}=\m^4 \int d^4x\,\Tilde{\de}^2\Tilde{\D}^2\fracmm{1}{2\p i}
\left[ \l\oint_{C_1}d\x\,\fracmm{Q^2}{2\x}+ \m^{-2} 
\oint_{C_2}d\x\,Q\ln Q\right]~,\eqno(5.11)$$
where the contour $C_1$ goes around the origin while the contour $C_2$ encircles
two roots of the quadratic equation $Q(\x)=0$ in the complex $\x$-plane.

In the massless limit $m\to 0$ one has $\l\to 0$ too, so that the action
$S_{\rm dual}$ is reduced to the improved N=2 tensor multiplet action alone. The
latter is N=2 superconformally invariant, and is equivalent to a {\it free} 
action~\cite{holl}, as it should have been expected for the massless 
hypermultiplet LEEA. It agrees with the statements made in ref.~\cite{sw2} about
the absence of quantum corrections to the hyper-K\"ahler effective potential for
the massless hypermultiplets, and the N=1 superspace calculations of 
ref.~\cite{gru} as well.

\section{Conclusion}

The N=2 extended supersymmetry severely restricts the effective hyper-K\"ahler
potential in the LEEA for N=2 matter fields. That becomes apparent in the N=2 
HSS where the N=2 matter is described by off-shell hypermultiplets, and the 
automorphism invariance $SU(2)_{\rm A}$ of the N=2 supersymmetry algebra 
restricts the effective hypermultiplet self-couplings even further. As a result,
the exact form of the effective hyper-K\"ahler potential is determined by a few
parameters to be related to a symmetry breaking in quantum theory. In the 
simplest case of N=2 QED with a single massive hypermultiplet, one finds a unique
solution given by the Taub-NUT metric. Any non-trivial solution thus represents
a counter-example to the naive non-renormalization `theorem', that can be tested 
in one-loop perturbation theory. There exists a manifestly N=2 supersymmetric
duality transformation which converts the hypermultiplet LEEA in the N=2 HSS into
the equivalent action in terms of an N=2 tensor multiplet. The duality 
transformation in that case also gives a simple root from the N=2 HSS to the 
components -- it is highly non-trivial in a generic situation.
 
I assumed that {\it non-analytic} effective  hypermultiplet self-couplings (they
are similar to that in eq.~(1.2), but in the N=2 HSS) do {\it not} contribute to
the low-energy effective hyper-K\"ahler potential. Though being quite natural by 
dimensional reasons, that assumption has yet to be justified in the non-abelian 
case.

\end{document}